\documentclass[fleqn,usenatbib]{mnras}
\usepackage{newtxtext,newtxmath}
\usepackage[T1]{fontenc}
\usepackage[dvips]{graphicx}
\usepackage{amsmath}
\usepackage{amsfonts}
\usepackage{comment}
\usepackage[dvipsnames, table]{xcolor}
\usepackage{hyperref}
\hypersetup{
	colorlinks=true,	% false: boxed links; true: colored links
	urlcolor=MidnightBlue,		% color of external links
	hypertexnames=true,
	plainpages=false,
	naturalnames=true,
	linkcolor=WildStrawberry,          % color of internal links (change box color with linkbordercolor)
	citecolor=ForestGreen,        % color of links to bibliography
}
\usepackage[normalem]{ulem}
\usepackage{supertabular}

\newcommand{\kepler}{{\it Kepler}}

\newcommand{\kepb}{Kepler-1708 b}
\newcommand{\kepbi}{Kepler-1708 b-i}

\newcommand{\multi}{{\sc MultiNest}}
\newcommand{\luna}{{\tt LUNA}}
\newcommand{\pandexo}{{\sc PandExo}}
\newcommand{\pandeia}{{\sc Pandeia}}

\newcommand{\pdf}{\mathrm{Pr}}

\hypersetup{
    pdftitle={K1708 b-i is likely undetectable with HST},
	pdfauthor={Cassese and Kipping}
	}

\title[K1708 b-i is likely undetectable with HST]{
Kepler-1708 b-i is likely undetectable with HST
}

\author[Cassese \& Kipping]{Ben Cassese$^{1}$\thanks{E-mail:
\href{mailto:b.c.cassese@columbia.edu}{b.c.cassese@columbia.edu}} and David Kipping$^{1}$\\
$^{1}$Dept. of Astronomy, Columbia University, 550 W 120th Street, New York NY 10027}

\date{Accepted July 20, 2022}%Accepted . Received ; in original form }

\pubyear{2022}

\begin{document}
\label{firstpage}
\pagerange{\pageref{firstpage}--\pageref{lastpage}}
\maketitle
%%%%%%%%%%%%%%%%%%% TITLE PAGE %%%%%%%%%%%%%%%%%%%

\begin{abstract}
The exomoon candidate \kepbi{} was recently reported using two transits of \kepler{} data. Supported by a 1\% false-positive probability, the candidate is promising but requires follow-up observations to confirm/reject its validity. In this paper, we consider the detectability of the exomoon candidate's transit, most specifically in the next window (March 2023) using the WFC3 instrument aboard the Hubble Space Telescope (HST). Using realistic noise estimates, accounting for the visit-long trends, and propagating the model posteriors derived using the \kepler{} data, we perform 75 injection-recovery trials with Bayesian model selection. Defining a successful detection as one which meets thresholds of the Bayes factor, AIC, and error of the retrieved parameters, only 7 of our 75 injections were recovered when considering HST data alone. This implies a true-positive probability (TPP) of $10\pm3$\%. Despite HST's superior aperture to \kepler{}, both instrumental systematics and the compactness of the candidate exomoon's orbit typically obfuscate a strong detection. Although the noise properties of the James Webb Space Telescope (JWST) have not yet been characterized in flight, we estimate the signal would be easily recovered using NIRSpec operating in its Bright Object Time Series mode.
\end{abstract}


\section{Introduction}
\label{sec:introduction}
After efforts ongoing for nearly as long as the hunt for exoplanets \citep{sartorettiDetectionSatellitesExtrasolar1999}, the search for their natural satellites (exomoons) has borne fruit in the most recent years. Two promising candidates have emerged: Kepler-1625 b-i \citep{teacheyEvidenceLargeExomoon2018} and \kepbi{} \citep{kippingExomoonSurvey702022}. However, should either of these moons truly exist, they do not resemble any satellites in our own solar system and each require independent observational confirmation. Until then, the unfamiliar scale of these planet-moon systems, combined with the subtlety/complications of their detected signals, motivates their ``candidate" rather than ``confirmed" statuses.

The unfamiliarity alone is not a reason to be skeptical of their existence, since hindsight will likely show that the first exomoons discovered were simply the easiest to detect, not representatives of the true population. Reports of the first transiting exoplanets followed this same discovery sequence, and only now is it clear that the earliest discovered Hot Jupiters (e.g. \citet{brownHubbleSpaceTelescopeTimeSeries2001}, \citet{konackiExtrasolarPlanetThat2003}) are not typical. However, with reported radii of 4.05 and 2.61 R$_\oplus$ respectively, the sizes of these systems defy most current models of moon formation and are difficult to explain.

Models which describe in-situ satellite growth in a circumplanetary disk place an upper limit on the size of a moon embryo: if a moonlet grows too large, disk torque interactions will initiate orbital decay and the satellite will either be lost via collision with the planet \citep{canupCommonMassScaling2006} or simply forced outside of its feeding zone \citep{batyginFormationGiantPlanet2020}. In a different scenario, moons can instead form elsewhere and experience subsequent capture by a giant planet. However, this mechanism requires several tuned/compatible factors, then even once captured, many systems will fail to circularize the initial loosely bound orbit and moon will be lost again \citep{porterPostCaptureEvolutionPotentially2011}.

Following the report of Kepler-1625 b-i, new models tailored specifically to that purportedly gigantic candidate claimed to support both in-situ formation and capture scenarios \citep{moraesExploringFormationScenarios2020a, hansenFormationExoplanetarySatellites}. However, their robustness cannot be established without comparisons to other exomoon candidates. For now, both confirmation of these strange moons and verification of the models which could potentially explain their existence rely on future observations with powerful observational facilities.

Although both candidates were first flagged via analysis of \kepler{} data \citep{teacheyHEKVIDearth2017, kippingExomoonSurvey702022}, Kepler-1625 b-i has already enjoyed follow up examination with HST \citep{teacheyEvidenceLargeExomoon2018}. These WFC3 observations added weight to the moon interpretation, but did not cleanly settle the issue due to the observations not covering the moon's egress and uncertainty in removal of transit-long systematics. Further analyses of the same data both recovered the moon signal \citep{hellerAlternativeInterpretationExomoon2019} and questioned it \citep{kreidbergNoEvidenceLunar2019}, leaving the status of the candidate unresolved.

\kepbi{} by contrast is the newest member of the catalog of potential exomoons and has not yet been observed by any facility besides \kepler{}. At first pass, the natural next step for confirming and potentially characterizing this candidate would be to observe the next transit of \kepb{} with a more powerful observatory. Since this hypothetical attempted recovery would rely on sub-ppt precision time series photometry, HST is the singularly capable facility for the job (prior to JWST's science operations). This impression of HST's capability grows stronger when comparing Kepler-1708 b and Kepler-1625 b: their host magnitudes ({\it J}=14.43, {\it J}=14.36, respectively) and transit durations (19.13 and 18.82 hours, respectively) are very similar, so presumably a similar observation as \citet{teacheyEvidenceLargeExomoon2018} modified to include a baseline long enough to catch moon ingress/egress should be decisive. However, we found that due to the smaller size of the moon and the compactness of its orbit, the same systematics which complicated analysis of Kepler-1625 b-i would likely doom any attempted recovery of \kepbi{}.

To reach this conclusion, we first considered the likely transit geometries of the system in the upcoming epoch, then created synthetic WFC3 light curves of the \kepb{} system under the assumption the moon exists. We fit these light curves with two models, one which considered a planet and a moon, and one which considered only a planet, then compared summary statistics comparing the models to determine how likely we were to correctly disentangle the moon from the planet. Justification for each step of this procedure and discussion about which known systematics were left out of our artificial observations follow below.

\section{Transit Geometry}
\label{sec:geo}
\subsection{Chance of Transit}
\label{sub:transit_odds}

The geometry of a planet-moon transit is more complex than its lone-planet counterpart, and unfortunately, is also more complex than just two superimposed single planetary systems. This is because the moon orbits its host on timescales shorter than the planet orbits its star, so at every transit epoch, the moon effectively has a different impact parameter and time of mid-transit. The situation gets worse if the moon's period is comparable with the planet's transit duration, since in that case the moon does not transit across the face of the star in a straight chord with a constant impact parameter. Algorithms such as \luna{} \citep{kippingLunaAlgorithmGenerating2011} handle this by numerically solving for the moon's position along a Keplerian orbit, so generating accurate light curves is still a tractable problem. However, differences in the shape of the light curve each epoch makes phase-folding light curves difficult \citep{kippingTransitOrigamiMethod2021} and observation planning more involved.

When specifically considering planning an observation of an exomoon transit, we have to consider two possible consequences of the moon's phase. First, if the moon's orbit around the planet is highly inclined and its orbital separation is comparable to the physical radius of the host star, there is a chance that the moon will dodge above or below the host star entirely as the plant transits. Although the moon's gravitational influence could still affect the timing of the planet's transit, this configuration would produce no observable dip in the combined light curve. \cite{martinTransitsInclinedExomoons2019} addressed the potential of this unfortunate alignment with an elegant analytic approximation which converts mutual inclination, the planet's impact parameter, and the moon's orbital radius into the chance of a misaligned transit by averaging over each possible moon phase.

\begin{figure}
\begin{center}
\includegraphics[width=8.4cm,angle=0,clip=true]{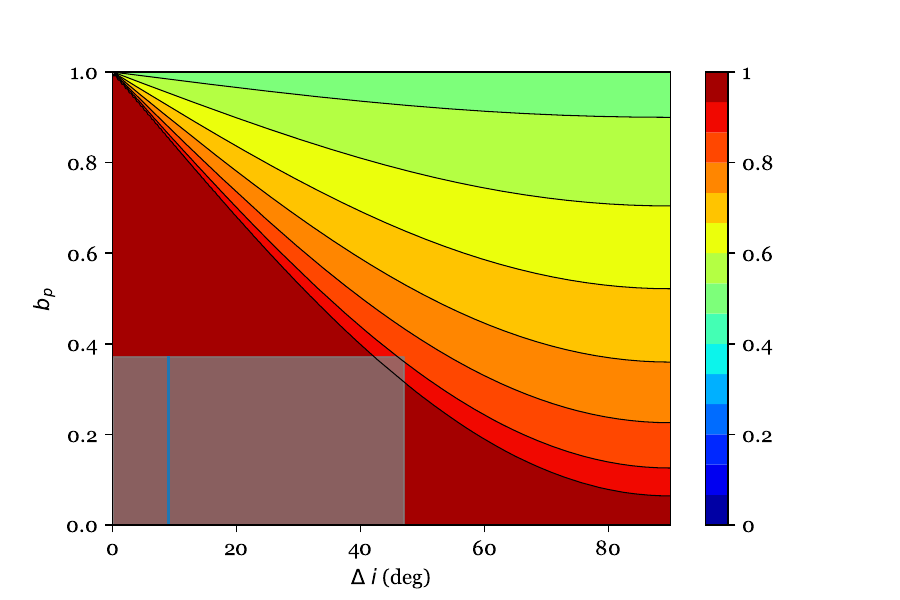}
\caption{
The likelihood that \kepbi{} transits when \kepb{} does. The grey box marks the region of parameter space favored in \citet{kippingExomoonSurvey702022}. Its height is defined by the reported 2$\sigma$ bound $b<0.37$ and its width represents the reported $9^{+38}_{-45}$ $^\circ$ inclination. The solid blue line marks the reported 9$^\circ$ inclination value subject to the impact parameter bound. The moon-planet separation is fixed at 11.7 planetary radii (0.957 stellar radii), again from \citet{kippingExomoonSurvey702022}. Although \citet{martinTransitsInclinedExomoons2019} offer a correction term for short-period moons to address the assumption that the phase remains fixed through the duration of a transit, we did not include it here since it only raises the odds of a transit, which is already nearly guaranteed.}
\label{fig:hide_and_seek}
\end{center}
\end{figure}

In Figure \ref{fig:hide_and_seek}, we applied this function to a range of mutual inclinations and planet impact parameters while holding the moon's orbital radius fixed at the nominal value reported in \citet{kippingExomoonSurvey702022} of 11.7 planetary radii (0.957 stellar radii). We see that since the moon's orbit is likely smaller than the star's radius, it is nearly guaranteed to transit every time \kepb{} does.

That is mostly good news for observation planning and is not always the case: as \cite{martinTransitsInclinedExomoons2019} show, the nominal parameters for Kepler-1625 b-i, should it exist, suggest that it will only transit 40\% of the times Kepler-1625 b does. However, the moon's transit signal is most easily resolved when its transit dip is well separated from the planet in time (i.e. it significantly leads or lags behind the planet on-sky), and this becomes less likely as its orbit shrinks. Kepler-1625 b-i's larger separation of 2.18 stellar radii \citep{teacheyEvidenceLargeExomoon2018} suggest that while its transits are uncommon, when they do occur, they should be well-separated in the light curve.

\subsection{Chance of Syzygy}
\label{sub:syzygy}

The second consequence we must consider is that the moon might ``hide" in front of or behind the planet. We refer to times with partial overlap between all three objects (star, planet, and moon) as a syzygy, and although \luna{} accounts for them when creating light curves, it does not report when they occur. Instead, we can check if a syzygy occurs for a given orbital geometry by numerically solving for the moon and planet position in time and checking if their centers ever fall within $R_p + R_m$. If we repeat this over a grid of possible initial moon phases, we can estimate chance of a syzygy for a given orbital configuration.

First, we can approximate the planet's $\{x,y\}$ position on the sky as $\mathcal{P}(t)$ via:
\begin{equation}
    \mathcal{P}(t) = \left\{\left(\frac{2 t}{d} - 1 \right) \sqrt{(1+R_p)^2-b^2} , b\right\}
\label{eq:planet}
\end{equation}
where $b$ is the planet's impact parameter, $d$ is the duration between first and last contact points, and $R_p$ is the planet radius.

Similar to \cite{martinTransitsInclinedExomoons2019}, the moon's orbit centered on the planet is:
\begin{equation}
\begin{split}
    \mathcal{O}_x(t) = a_m \left(\cos(\Omega) \cos(f(t)) - \cos(i) \sin(\Omega) \sin(f(t)) \right)
    \\
    \mathcal{O}_y(t) = a_m \left(\sin(\Omega) \cos(f(t)) + \cos(i) \cos(\Omega) \sin(f(t)) \right)
\end{split}
\label{eq:orbit}
\end{equation}
Where $\Omega$ is the moon's longitude of ascending node, $i$ is the moon's inclination relative to the planet's, and $f(t)$ is the moon's true anomaly as a function of time. This we can calculate as $f(t) = \phi_i + (2 \pi/P_m) t$, where $\phi_i$ is the arbitrary ``initial" phase of the moon and $P_m$ is the period of the moon. Combining Equations \ref{eq:planet} and \ref{eq:orbit}, we get the moon's position $\mathcal{M}(t) = \mathcal{P}(t) + \mathcal{O}(t).$ To determine if a set of parameters will result in a syzygy, we calculate:
\begin{equation}
    \begin{cases}
    \text{Full syzygy occurs} &  \min(|\mathcal{P}(t) - \mathcal{M}(t)|) <= R_p - R_m \\
    \text{Partial syzygy} &  R_p - R_m < \min(|\mathcal{P}(t) - \mathcal{M}(t)|) <= R_p + R_m
    \\
    \text{No syzygy occurs} & R_p + R_m < \min(|\mathcal{P}(t) - \mathcal{M}(t)|)
    \end{cases}
\end{equation}
for $0<=t<=d$.

\begin{figure}
\begin{center}
\includegraphics[width=8.4cm,angle=0,clip=true]{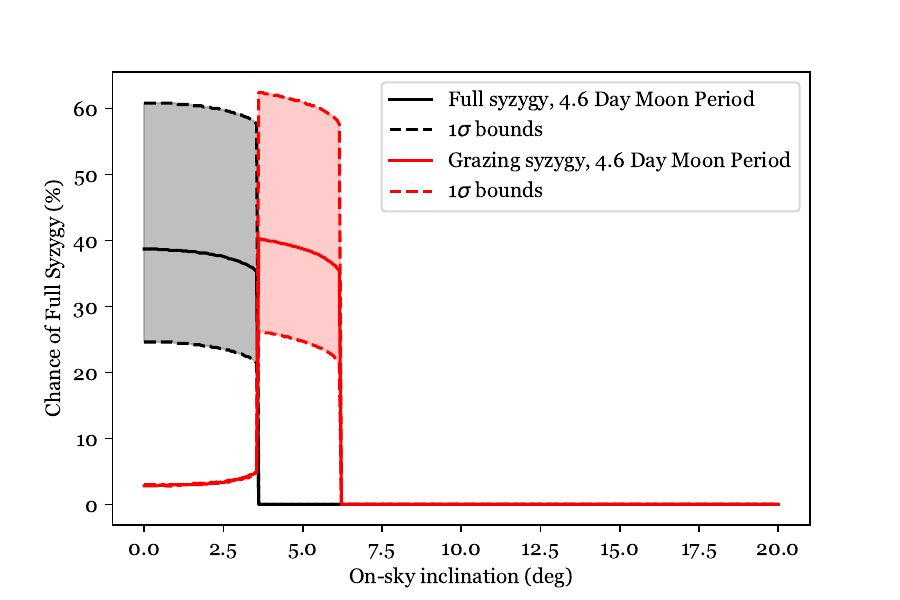}
\caption{The chance of a syzygy as a function of moon-planet inclination. The solid black line shows the chance of a full syzygy in which the moon is completely hidden assuming the nominal 4.6 day moon period, and the solid red line shows the same for grazing-only syzygies. The dashed lines show the 1$\sigma$ bounds in moon period. Note that the reported mutual inclination was $9^{+38}_{-45}$ $^{\circ}$, so the majority of the likely phase space will result in no syzygys, but a small slice of it can result in up to 60\% chance. For further context, the Galilean satellites all have inclinations of $<3^\circ$, although Titan has an inclination of $27^\circ$. Moons in the Solar System are typically aligned to their planet's equators, not their planet's orbital planes, so the exomoon inclination distribution likely tracks the exoplanet obliquity distribution.}
\label{fig:syzygy}
\end{center}
\end{figure}

Since the chance of a syzygy depends only on the separation between the planet and the moon, and this does not depend on $\Omega$ or $b$, we can arbitrarily set those to zero and focus on the affects of $i$ and $a_m$\footnote{However, if we wanted to calculate the chance of a grazing moon transit across the star or the path the moon takes across the disk (removing the assumption it remains fixed in its orbit during the transit), we would have to include these parameters}. This is shown in Fig \ref{fig:syzygy}, which demonstrates that at $|i|<4^\circ$, a full syzygy is possible. The fits to \kepler{} data favor an inclination of  $i = 9^{+38}_{-45}$ $^\circ$, so syzygies are possible but not necessarily preferred.

Lastly, we note that although calculating the chance of a syzygy requires numerically solving the moon's motion, finding the critical inclination values at which syzygies become possible does not. Calculating the required inclination to allow for a syzygy is analogous to calculating the required inclination for an isolated planet to transit across a star, and is a function of separation. The maximum inclinations on-sky the moon's orbit can have while leaving open the possibility of a syzygy is:

\begin{equation}
    \begin{cases}
    \text{Full syzygy possible} & |i| < \pi/2-\arccos{(R_p - R_m)/a_m}
    \\
    \text{Partial syzygy possible} & |i| < \pi/2-\arccos{(R_p + R_m)/a_m}
    \end{cases}
\end{equation}

Using the nominal values from \cite{kippingExomoonSurvey702022}, the on sky inclination must be $i<3.6^\circ$ for a full syzygy and $i<6.2^\circ$ for a grazing syzygy.

\subsection{Expected Moon Phase}
\label{sub:expected_phase}

Both our calculations for the likelihood of a non-transit (Section \ref{sub:transit_odds}) and of a syzygy (Section \ref{sub:syzygy}) averaged over all possible moon phases, meaning we implicitly assumed there is no preferred phase at the time of a hypothetical observation. This initially seems in tension with our stated method of sampling from the posterior distributions of parameters fit to the original \kepler{} data: since the marginalized posterior for moon phase is not uniform, one might think we have some predictive ability when considering the phase at upcoming epochs. However, this is unfortunately not the case, since the posterior distributions for planet period and, more importantly, moon period both allow for uncertainty.

\begin{figure*}
\begin{center}
\includegraphics[width=17.0cm,angle=0,clip=true]{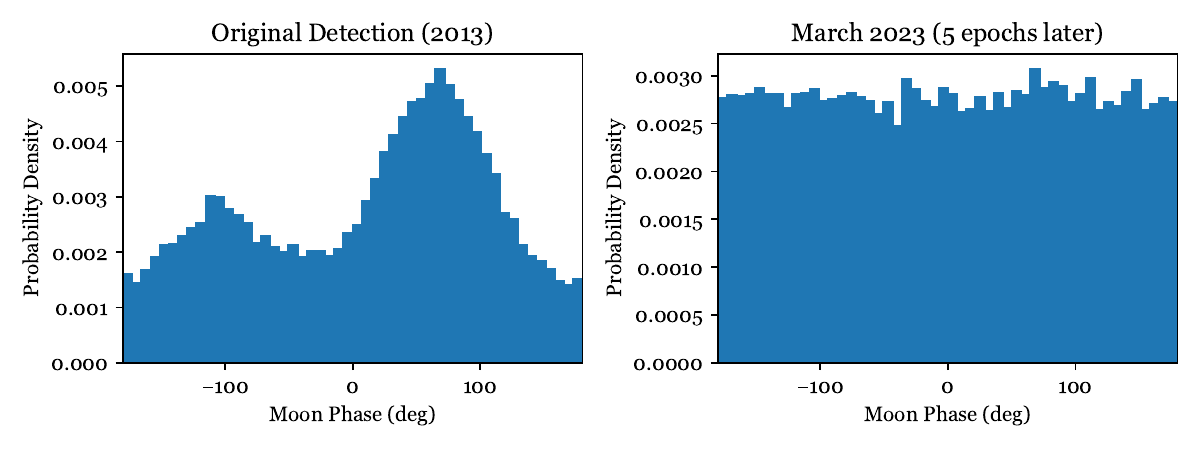}
\caption{The phase (true anomaly) of the moon. The left panel shows the marginalized posterior distribution of the phase as fit to the original \kepler{} data, and the right shows the posterior distribution propagated forward by five planet orbits. Since the period of the moon's orbit is uncertain, the resulting phase distribution flattens over time as we lose predictive power.}
\label{fig:phases}
\end{center}
\end{figure*}

Although the planet's period is constrained to $\sim$minutes, the moon's orbital period has a nearly factor of 2 upper uncertainty at $4.6\substack{+3.1 \\ -1.8}$\, days. Since it has been about a decade since the last observation of a transit of \kepb{}, this uncertainty in the moon's period washes out any serious chance of predicting the moon's orbital phase for any upcoming transits, as shown in Fig \ref{fig:phases}. Although we have focused this study on the next transit occurring in March of 2023, there is no reason to think this transit will be more or less favorable for detection than any other.

Having established that the moon is likely to transit the star each epoch, reside close to the planet on-sky (and therefore convolved in the light curve), and has a chance of aligning with the planet in a syzygy, we now turn to simulating future observations.

\section{Injection-Recovery}
\label{sec:post}
\subsection{Predicting HST's Photometric Performance}
\label{sub:noise}

The first step in creating our injected light curves was determining the sampling cadence. For this we used the HST module of \pandexo{} \citep{batalhaPandExoCommunityTool2017}, a widely used Python package mainly dedicated to predicting noise in JWST spectral measurements. We selected instrument settings which resulted in the lowest total uncertainty on transit depth, which collectively resulted in 5 minute exposures with the G141 grism. These were the same settings used in the actual observations of Kepler-1625 b by \citet{teacheyEvidenceLargeExomoon2018}.

Next, we needed to derive estimates of the RMS noise on each point in the light curve. \pandexo{} also supplies this, but we noted that when simulating observations of the Kepler-1625 b system, it returned an optimistic value of 339\,ppm. This is much lower than the actually measured value in  \citet{teacheyEvidenceLargeExomoon2018}, which was either 375.5 or 440.1\,ppm depending on choice of ramp/hook removal technique (discussed below). While \pandexo{}'s JWST noise estimations rely in part on detector-level estimates from the \pandeia{} instrument simulator \citep{pontoppidanPandeiaMultimissionExposure2016}, its treatment of noise on HST estimates instead rely on scaling from a standard source. Switching that source from the default measurements of the GJ 1214 b system to values measured for Kepler-1625 b resulted in estimates of either 387 or 453\,ppm, again depending on the ramp/hook removal technique. For the purpose of demonstrating that a non-detection is the most likely outcome, we used a slightly optimistic value of 381\,ppm.

In addition to this Gaussian noise behaviour, WFC3 light curves are known to exhibit three prominent systematic noise effects \citep{wakefordMarginalisingInstrumentSystematics2016}: breathing, ramp/hook, and visit-long trend. Breathing tracks with the orbital phase of HST and thus can be easily removed when the astrophysical timescale of interest is not commensurate (such as the case of for \kepb{}'s 19\,hour transit). The ramp/hook also occurs on this timescale, caused by charge trapping within the detector \citep{wakefordMarginalisingInstrumentSystematics2016}. For transits with much longer durations, such as \kepb{}, the hooks can be effectively removed using a data-driven non-parametric method introduced in \citet{teacheyEvidenceLargeExomoon2018} (this method resulted in the 375.5\,ppm estimate of the RMS, while fitting an exponential model to the hook resulted in 440.1\,ppm). We thus expect that these two effects can be removed at a level below the photon noise in what follows. The same cannot be said of the visit-long trend and this will be discussed in the next subsection.

\subsection{Injection}
\label{sub:injection}

Having determined realistic Gaussian noise and sampling cadence in Section~\ref{sub:noise}, we next proceeded to create the artificial light curves. To accomplish this, we started with 75 randomly selected samples of the joint posterior of the planet+moon model fit from \citet{kippingExomoonSurvey702022}. We then used \luna{} \citep{kippingLunaAlgorithmGenerating2011} to simulate the expected ``true" transit signal each of these parameter vectors would produce for the epoch occurring on March 24th, 2023. These 75 light curves were then sampled and noised using the results from Section~\ref{sub:noise}.

At this stage, we returned to systematic effects. As noted in Section~\ref{sub:noise}, a systematic that is likely to be covariant with the planet+moon model parameters is the visit-long trend. Occurring on the timescale an entire visit rather than a single orbit, this is commensurate with the transit duration and was found in \citet{teacheyEvidenceLargeExomoon2018} to have a strong impact on the inferred moon transit depth. Several trend models have been proposed in the literature, including a linear slope \citep{huitsonHSTOpticaltonearIRTransmission2013,fraineWaterVapourAbsorption2014,ranjanAtmosphericCharacterizationFive2014,knutsonHubbleSpaceTelescope2014}, a quadratic \citep{stevensonHubbleSpaceTelescope2014, stevensonTransmissionSpectroscopyHot2014} and an exponential \citep{teacheyEvidenceLargeExomoon2018}, but here we adopt the quadratic model as a flexible but simple approach.

Unfortunately, there is a final complication to consider before applying this quadratic to the entire transit. \kepbi{} has a long transit duration of $(19.13\pm0.19)$\,hours \citep{kippingExomoonSurvey702022}, which means that covering this timescale alone requires 12 HST orbits. To include some out-of-transit baseline, which is necessary both to establish the out-of-transit flux and to search for the moon ingress/egress, we expect that at least two visits would be required to accomplish the observations\footnote{STSci recommends breaking observations of \href{https://www.stsci.edu/hst/proposing/phase-ii/visit-size-recommendations}{>5 orbits into multiple visits}, though the 26 orbits in \citet{teacheyEvidenceLargeExomoon2018} were broken into 2 visits.}. An unavoidable consequence of visit changes is that the target lands on a slightly different position on the detector after each movement, which introduces a flux offset due to inter pixel response variations \citep{teacheyEvidenceLargeExomoon2018}. This flux offset is crucial to include in our injection because it occurs inside the planetary transit and thus resembles an exomoon ingress/egress feature \citep{kippingLunaAlgorithmGenerating2011}. Therefore, to correctly include the visit-long trend, we must actually apply two quadratics (one for each visit) separated by an offset. Accordingly, for each injected light curve, we multiply the noised, sampled planet+moon simulation with a systematics model given by

\begin{equation}
S(t) =
\begin{cases}
a_0 + a_1 (t-t_{\mathrm{ref}}) + a_2 (t-t_{\mathrm{ref}})^2 & \text{if } t < t_{\mathrm{ref}} ,\\
b_0 + a_1 (t-t_{\mathrm{ref}}) + a_2 (t-t_{\mathrm{ref}})^2 & \text{if } t > t_{\mathrm{ref}} ,
\end{cases}
\end{equation}

where $a_0$, $b_0$, $a_1$ and $a_2$ are parameters defining the visit-long trend, $(a_0-b_0)$ defines the visit-change flux offset, and $t_{\mathrm{ref}}$ is chosen to equal to the expected (maximum likelihood) mid-transit time for \kepb{}. For the actual choice of these parameters, we again used 75 random rows from the joint planet+moon model posterior in \citet{teacheyEvidenceLargeExomoon2018}.

\subsection{Recovery}
\label{sub:recovery}

Equipped with 75 realistic simulations of the breathing+ramp/hook corrected HST WFC3 observations of \kepb{} and its candidate companion, we were now ready to attempt to recover the moon. In this recovery exercise, we conceived that the goal was to independently detect \kepbi{} using new data alone. Accordingly, we treat the HST light curve in isolation in what follows, rather than jointly fitting the HST+\kepler{} data together.

For each simulated light curve, we regressed two different models using \multi{} \citep{ferozMultiNestEfficientRobust2009}, a nested sampling algorithm which naturally computes the Bayesian evidence of a given model fit. The first model describes a planet+moon system with the same 14 transit parameters describing the system as used in \citet{kippingExomoonSurvey702022} plus the additional 4 parameters needed to describe HST's visit-long trend ($a_0, a_1, a_2$, and $b_0$), giving 18 in total. The second model describes a planet-only system and uses the same parameters, minus those which describe the moon. To be considered a successful recovery, the planet+moon model must both correctly describe the injected signal and also offer enough of an improvement over the planet-only model to justify the inclusion of the additional parameters. We quantify this improvement by declaring a successful recovery must meet thresholds on the Bayes factor, Akaike Information Criterion (AIC), and $\chi^2$ as described below.

Beginning with the Bayes Factor, we follow \citet{kippingExomoonSurvey702022} and use the \citet{kassBayesFactors1995} table to define a ``strong detection'' as that with a Bayes factor exceeding 10. Using this measure alone, 29 of our 75 experiments could be classified as successful recoveries. However, when examining these recovered signals, we found that many of the maximum a posteriori (MAP) solutions did not resemble the true injected light curve, and further that many of these featured no visible moon signal at all. Instead, many fits took advantage of the hide-and-seek misalignment described in Section \ref{sub:transit_odds} and placed its MAP moon on a wide orbit in a non-transiting geometry (Table \ref{table:sims}). While many of these suspicious ``detections" can be rejected through examination of their suggested parameters (nearly all favored moons with unphysical densities, for instance), their ability to meet this Bayes factor threshold requires a stricter detection criterion.

The breakdown of the Bayes factor's ability to cleanly differentiate successes and failures necessitated the inclusion of additional criteria, and in general should illustrate the caution required when fitting single transits of planet+moon systems. In principle, these configurations would be disfavored if we considered more than one epoch of transit data, since such alignments will not repeat (excluding the truly unfortunate scenario in which the planet and moon periods are in an integer ratio). But, by restricting ourselves to this one-transit case, there is a large region of planet+moon parameter space which can describe the general shape of a planet transit very well.

With that in mind, we mandated that each successful trial must not only yield a Bayes factor of $>10$, they must also have a difference in $\chi^2 > 9$ and a difference in AIC $<0$. These conditions, when considered individually paired with the Bayes factor condition, both return the same subset of trials. Finally, we additionally required that each successful trial had to also capture the true injected moon radius and separation within its $2\sigma$ marginalized posterior. All told, 29 trials passed the Bayes factor threshold, 27 met the $\Delta \chi^2$ bar, 21 satisfied $\Delta$AIC$<0$, and 7 successfully recovered the injected $R_m$ and $a_m$. The union of these thresholds includes 7 trials, and moving forward we call these our successful recoveries. Statistics for all trials are included in Table \ref{table:sims}.

The priors on the 14 transit parameters were identical to that used in \citet{kippingExomoonSurvey702022}, with the exception of orbital period and transit mid-time. Similarly, the priors on the 4 visit-long trend parameters are identical to that of \citet{teacheyEvidenceLargeExomoon2018}. For the planetary period and mid-transit time of \kepb{}, we used informative priors since the expected time of transit would be not arbitrary but instead fall within a narrow range governed by the previous \kepler{} data. We thus take the posterior distribution for the period and transit time from the planet+moon model of \kepb{} reported in \citet{kippingExomoonSurvey702022} and use it to define an informative prior. We find that two independent normals well describe the resulting posteriors with $P=(737.1094\pm0.0032)$\,days
and $\tau=(2455605.1849\pm0.0021)$\,BJD.

\begin{figure*}
\begin{center}
\includegraphics[width=17.0cm,angle=0,clip=true]{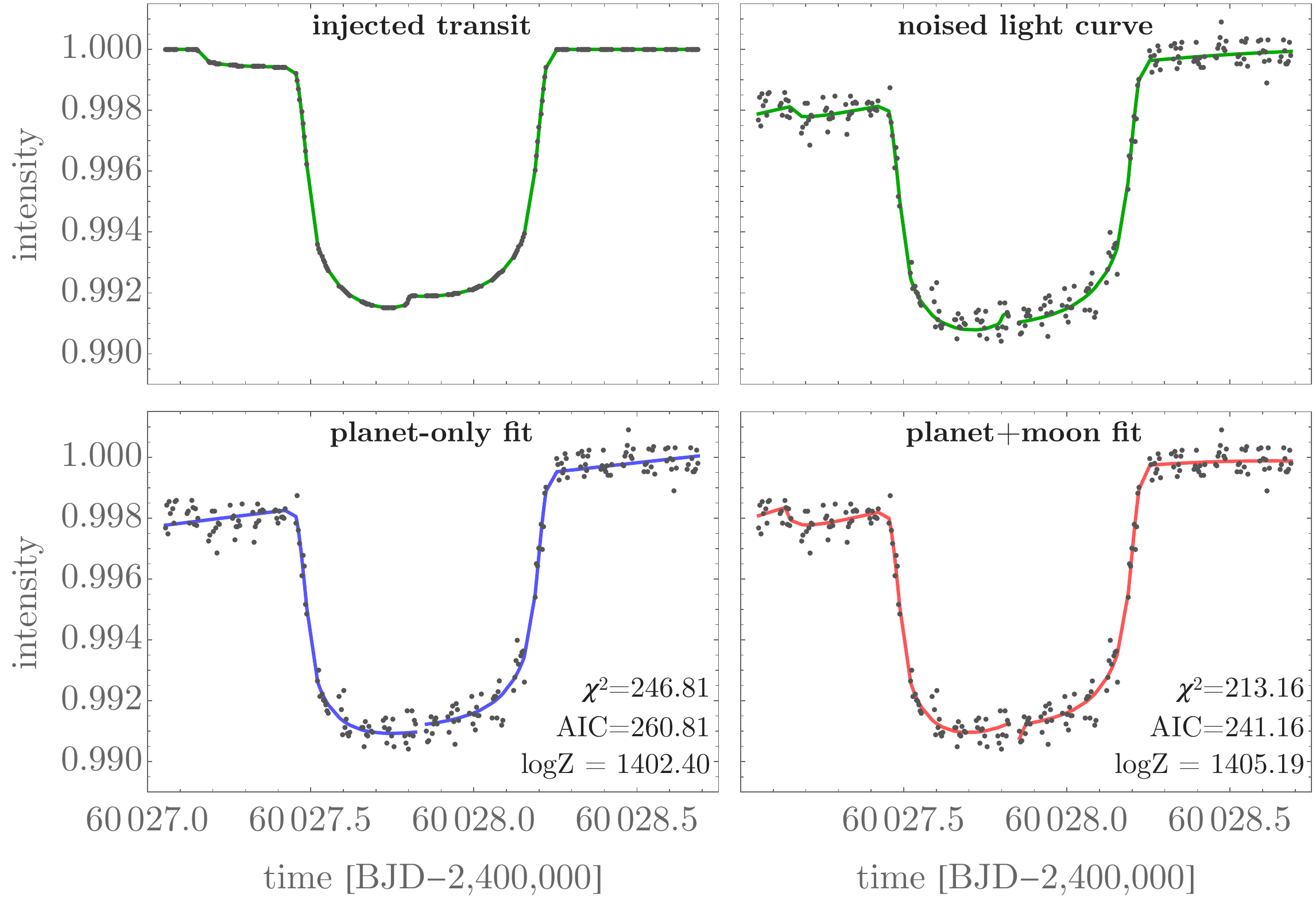}
\caption{
The full injection-recovery process, showed for simulation \#13. Top-left shows the predicted light curve of \kepb{} and \kepbi{} in the next transit window (green solid), in March 2023, sampled at the cadence expected using HST (black points). Top-right shows the same but after injecting a realistic visit-long trend systematic and adding WFC3 noise. Bottom panels show the planet-only fit (left) and planet+moon fit (right). In this case, the log Bayes factor was $2.79$, the difference in $\chi^2$ was 33.65, the difference in AIC was -19.65, and the ``true" injected moon radius and separation fell well within the $2\sigma$ bounds of the marginalized posterior. Thus, this is considered a successful detection.}
\label{fig:lc}
\end{center}
\end{figure*}

The entire end-to-end process of signal simulation, noise injection and
successful recovery is depicted for simulation \#13 as an example in
Figure~\ref{fig:lc}.

\subsection{True Positive Probability}
\label{sub:TPP}

\begin{figure}
\begin{center}
\includegraphics[width=8.4cm,angle=0,clip=true]{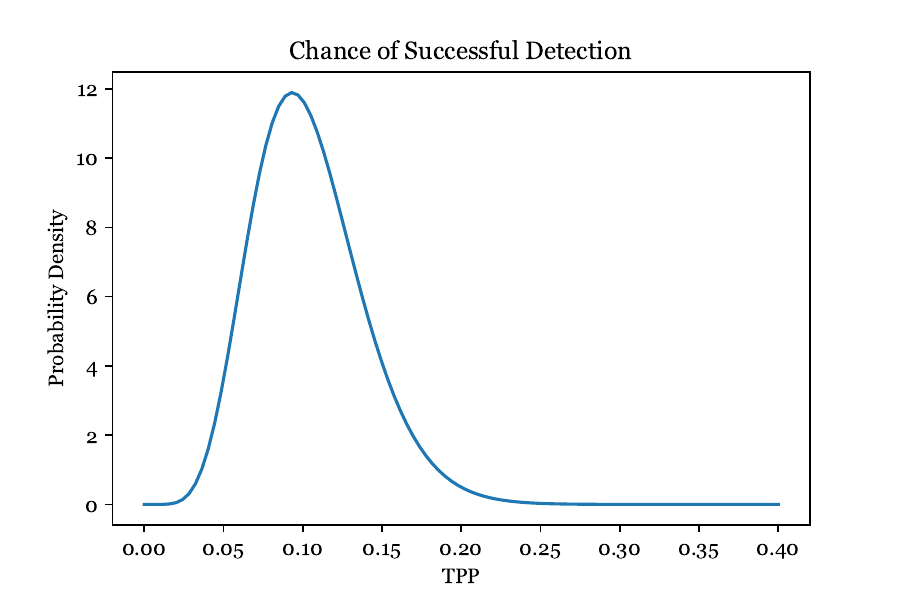}
\caption{
The distribution of true positive probability that HST would recover \kepbi{} with a single epoch of data, should it exist. Analytically the expectation is $10 \pm 3$\%, and numerically the peak (mode) and $1\sigma$ bounds are $9.3^{+3.3}_{-3.5}$\%.
\citet{kippingExomoonSurvey702022}}
\label{fig:tpp}
\end{center}
\end{figure}

Since detection vs. non-detection is a binary problem and thus has an associated Bernoulli probability of success, we can analytically use our 7 successes in 75 trials to calculate the chances that a single real observation would recover the moon, should it exist. This chance is the true positive probability (TPP), also known as completeness. The probability of obtaining $M$ successes from $N$ Bernoulli experiments with a success rate of $\mathrm{TPP}$ defines a Binomial distribution, such that

\begin{align}
\pdf(M|N,\mathrm{TPP}) = \mathrm{TPP}^M  (1 - \mathrm{TPP})^{N-M} \binom{N}{M}.
\end{align}

If we wish to infer a posterior distribution for $\mathrm{TPP}$ conditioned upon $M=7$ successes and $N=75$ trials, then the above represents the likelihood function. The prior on $\mathrm{TPP}$ may be assumed to be uniform for simplicity, leading to a posterior of

\begin{align}
\pdf(\mathrm{TPP}|M,N) \, =& \, \frac{\pdf(M|N,\mathrm{TPP}) \pdf(\mathrm{TPP})}{ \int_{\mathrm{TPP}=0}^1 \pdf(M|N,\mathrm{TPP}) \pdf(\mathrm{TPP}) \,\mathrm{d}\mathrm{TPP}},\nonumber\\
=& \, (1+N) (1-\mathrm{TPP})^{N-M} \mathrm{TPP}^M \binom{N}{M}.
\end{align}

This function is plotted in the inset of Figure~\ref{fig:tpp}. The expectation value of $\mathrm{TPP}$, $\overline{\mathrm{TPP}}$, is given by

\begin{align}
\overline{\mathrm{TPP}} =& \, \int_{\mathrm{TPP}=0}^1 \mathrm{TPP} \times \pdf(\mathrm{TPP}|M,N) \,\mathrm{d}\mathrm{TPP},\nonumber\\
=& \, \frac{M+1}{N+2},
\end{align}

and the variance by

\begin{align}
\sigma_{\mathrm{TPP}}^2 =& \int_{\mathrm{TPP}=0}^1 (\mathrm{TPP}-\overline{\mathrm{TPP}})^2 \pdf(\mathrm{TPP}|M,N) \,\mathrm{d}\mathrm{TPP},\nonumber\\
=& \frac{(M+1)(N-M+1)}{(N+2)^2(N+3)}. \label{eq:var}
\end{align}

Using $N=75$ and $M=7$, we can thus write that $\mathrm{TPP} = (0.10\pm0.03)$. On this basis, the most likely outcome of an HST follow-up effort is a failure to detect the exomoon signature of \kepbi{}.

\section{Discussion}
\label{sec:discussion}
%\subsection{JWST Moon Detection} % We cut the other subsection so now this is the whole discussion
%\label{sub:JWST_moon}

Although efforts to validate the existence of the satellite with HST would likely fail, a much more promising story unfolds if we consider instead using JWST. Now safely placed in its L2 orbit and into its commissioning phase, JWST and its NIRSpec instrument in particular promise to push exoplanet science into a new era defined by previously unobtainable data. The enhanced capabilities of JWST benefit exomoon hunting especially and will push many targets from the edge of detectability with HST into the realm of confident detections/non-detections.

We did not perform the same injection-recovery exercise with a simulated NIRSpec prism light curve as we did with HST WFC3 This is because such an exercise is not yet possible- we created our artificial HST light curves by applying a systematic model fit to existing data, and JWST has yet to collect any comparable data. Once its noise properties have been established for faint stellar targets, in principle the same injection-recovery procedure carried out for HST could confidently establish JWST's capabilities. 

However, even before considering these eventual JWST observations, its NIRSpec instrument offers four tempting advantages to an exomoon transit observation over HST's WFC3. First, the NIRSpec's low resolution prism mode provides spectral intensities between 0.6-5.3\,$\mu$m\footnote{\href{https://jwst-docs.stsci.edu/jwst-near-infrared-spectrograph/nirspec-operations/nirspec-bots-operations/nirspec-bots-wavelength-ranges-and-gaps}{JWST NIRSpec BOTS Operations}} as compared to WFC3's G141 grism's coverage between 1.075-1.7\,$\mu$m\footnote{\href{https://www.stsci.edu/hst/instrumentation/wfc3/documentation/grism-resources}{WFC3 Grism Resources}}. Star spot crossings produce wavelength dependent effects \citep{pontDetectionAtmosphericHaze2008} while transiting moons do not, so this increased spectral span can help distinguish between the two explanations in the case of moon egress during planet transit. Second, while HST must chunk long transits into multiple visits, JWST does not need to look away from its target and reacquire a guide star, even during high-gain antenna movements \citep{beanTransitingExoplanetCommunity2018}. This should mitigate target placement issues discussed in \ref{sub:injection}. Third, JWST's placement at L2 allows it to observe without interruptions caused by earth occultations, which allows for more time on target during a transit. Fourth, finally, and most obviously, JWST's mirror is >2.5x larger than HST's. This enormous difference in light gathering power will allow more precise observations at a far higher cadence and is key to resolving the subtle dip of a transiting moon.

We can add some quantitative rigor to these advantages even without performing full injection-recovery tests. Although there are not yet any in-flight observations of a faint source to determine NIRSpec's noise properties, several tools to simulate these observations are already available. Each of these each strongly imply that \kepbi{}, should it exist, is easily detectable. We rearranged the outputs of \pandexo{} \citep{batalhaPandExoCommunityTool2017}, which is primarily designed to predict spectral data, and \pandeia{} \citep{pontoppidanPandeiaMultimissionExposure2016}, which gives estimates of detector output for a single exposure (i.e. one integration, in the case of transit applications), into estimates of noise in a white light curve. Both tools predict that $\sim$12 second integrations of $\sim$50 groups each would each have a precision of 400-500\,ppm. The nominal parameters of the moon reported in \citet{kippingExomoonSurvey702022} would produce a transit dip of $\sim$450ppm, so over the >5,000 unbroken integrations covering the duration of the main transit the moon should easily stand out even with severe instrument systematics.

Excitingly, \cite{rustamkulovAnalysisJWSTNIRSpec2022} recently bolstered the hope that NIRSpec noise truly can be binned down in time without worry of a noise floor of comparable size to this measurement. They used data collected by NIRSpec pre-flight during cyro-vacuum testing to constrain the noise floor to <14 ppm, meaning we could bin upwards of three hours of data here before having to consider errors beyond shot noise. We finally note that these simulation tools also suggest that \kepbi{} is a challenging but potentially viable target for atmosphere retrieval. This is an especially exciting side-benefit worth further investigation given \kepb{}'s membership in the rare class of cool giant planet analogs.

Finally, we must acknowledge an unfortunate consequence of timing and temper some of our JWST enthusiasm. Although observations planning tools limit our temporal horizons to a few years due to uncertainties in the spacecraft's orbit (especially prior to launch), if patterns hold, they suggest that Kepler-1708 will not be visible for several more transit epochs. \kepb{}'s 737 day orbit is annoyingly close to exactly 2 years, meaning we must wait many epochs for transits to fall in a more favorable season. If current predictions prevail, \kepb{} will not be observable with JWST until at least its 2033 transit.

In summary, we estimate that even if \kepbi{} exists, there is only a $10 \pm 3$ \% chance that a single transit observation with HST would successfully detect it. This high chance of failure is driven by systematics which cannot easily be removed, such as the visit-long trend and the necessity to break a long transit into multiple visits. However, a single transit observation with JWST's NIRSpec instrument would likely recover the moon, should it be real, and could possibly provide the first characterization of a cool giant atmosphere, but likely only after many years due to transit visibility constraints. We look forward to future follow ups of this (and other) exciting targets in the upcoming era of exomoon science enabled by new instruments.

\section*{Acknowledgments}
\label{sec:acknowledgments}
This research has made use of the NASA Exoplanet Archive, which is operated by the California Institute of Technology, under contract with the National Aeronautics and Space Administration under the Exoplanet Exploration Program.

This research made use of Astropy,\footnote{\url{http://www.astropy.org}} a community-developed core Python package for Astronomy \citep{theastropycollaborationAstropyCommunityPython2013, theastropycollaborationAstropyProjectBuilding2018}.

BC thanks Soichiro Hattori for his helpful comments during preparation of the manuscript.

Special thanks to donors to the Cool Worlds Lab:
Mark Sloan,
Douglas Daughaday,
Andrew Jones,
Elena West,
Tristan Zajonc,
Chuck Wolfred,
Lasse Skov,
Graeme Benson,
Alex de Vaal,
Mark Elliott,
Methven Forbes,
Stephen Lee,
Zachary Danielson,
Chad Souter,
Marcus Gillette,
Tina Jeffcoat,
Jason Rockett,
Scott Hannum,
Tom Donkin,
Andrew Schoen,
Jacob Black,
Reza Ramezankhani,
Steven Marks,
Philip Masterson,
Gary Canterbury,
Nicholas Gebben,
Joseph Alexander \&
Mike Hedlund.

\section*{Data Availability}
Code and data produced during the preparation of this manuscript is available at \href{https://github.com/ben-cassese/Kep-1708b-i_Detectability}{this github repository}.

\bibliographystyle{mnras}
\bibliography{references}

\begin{thebibliography}{}
\makeatletter
\relax
\def\mn@urlcharsother{\let\do\@makeother \do\$\do\&\do\#\do\^\do\_\do\%\do\~}
\def\mn@doi{\begingroup\mn@urlcharsother \@ifnextchar [ {\mn@doi@}
  {\mn@doi@[]}}
\def\mn@doi@[#1]#2{\def\@tempa{#1}\ifx\@tempa\@empty \href
  {http://dx.doi.org/#2} {doi:#2}\else \href {http://dx.doi.org/#2} {#1}\fi
  \endgroup}
\def\mn@eprint#1#2{\mn@eprint@#1:#2::\@nil}
\def\mn@eprint@arXiv#1{\href {http://arxiv.org/abs/#1} {{\tt arXiv:#1}}}
\def\mn@eprint@dblp#1{\href {http://dblp.uni-trier.de/rec/bibtex/#1.xml}
  {dblp:#1}}
\def\mn@eprint@#1:#2:#3:#4\@nil{\def\@tempa {#1}\def\@tempb {#2}\def\@tempc
  {#3}\ifx \@tempc \@empty \let \@tempc \@tempb \let \@tempb \@tempa \fi \ifx
  \@tempb \@empty \def\@tempb {arXiv}\fi \@ifundefined
  {mn@eprint@\@tempb}{\@tempb:\@tempc}{\expandafter \expandafter \csname
  mn@eprint@\@tempb\endcsname \expandafter{\@tempc}}}

\bibitem[\protect\citeauthoryear{{Astropy Collaboration} et~al.,}{{Astropy
  Collaboration}
  et~al.}{2013}]{theastropycollaborationAstropyCommunityPython2013}
{Astropy Collaboration} et~al., 2013, \mn@doi [\aap]
  {10.1051/0004-6361/201322068}, \href
  {https://ui.adsabs.harvard.edu/abs/2013A&A...558A..33A} {558, A33}

\bibitem[\protect\citeauthoryear{{Astropy Collaboration} et~al.,}{{Astropy
  Collaboration}
  et~al.}{2018}]{theastropycollaborationAstropyProjectBuilding2018}
{Astropy Collaboration} et~al., 2018, \mn@doi [\aj] {10.3847/1538-3881/aabc4f},
  \href {https://ui.adsabs.harvard.edu/abs/2018AJ....156..123A} {156, 123}

\bibitem[\protect\citeauthoryear{Batalha et~al.,}{Batalha
  et~al.}{2017}]{batalhaPandExoCommunityTool2017}
Batalha N.~E.,  et~al., 2017, \mn@doi [Publications of the Astronomical Society
  of the Pacific] {10.1088/1538-3873/aa65b0}, 129, 064501

\bibitem[\protect\citeauthoryear{Batygin \& Morbidelli}{Batygin \&
  Morbidelli}{2020}]{batyginFormationGiantPlanet2020}
Batygin K.,  Morbidelli A.,  2020, \mn@doi [The Astrophysical Journal]
  {10.3847/1538-4357/ab8937}, 894, 143

\bibitem[\protect\citeauthoryear{Bean et~al.,}{Bean
  et~al.}{2018}]{beanTransitingExoplanetCommunity2018}
Bean J.~L.,  et~al., 2018, \mn@doi [Publications of the Astronomical Society of
  the Pacific] {10.1088/1538-3873/aadbf3}, 130, 114402

\bibitem[\protect\citeauthoryear{Brown, Charbonneau, Gilliland, Noyes  \&
  Burrows}{Brown et~al.}{2001}]{brownHubbleSpaceTelescopeTimeSeries2001}
Brown T.~M.,  Charbonneau D.,  Gilliland R.~L.,  Noyes R.~W.,   Burrows A.,
  2001, \mn@doi [The Astrophysical Journal] {10.1086/320580}, 552, 699

\bibitem[\protect\citeauthoryear{Canup \& Ward}{Canup \&
  Ward}{2006}]{canupCommonMassScaling2006}
Canup R.~M.,  Ward W.~R.,  2006, \mn@doi [Nature] {10.1038/nature04860}, 441,
  834

\bibitem[\protect\citeauthoryear{Feroz, Hobson  \& Bridges}{Feroz
  et~al.}{2009}]{ferozMultiNestEfficientRobust2009}
Feroz F.,  Hobson M.~P.,   Bridges M.,  2009, \mn@doi [Monthly Notices of the
  Royal Astronomical Society] {10.1111/j.1365-2966.2009.14548.x}, 398, 1601

\bibitem[\protect\citeauthoryear{Fraine et~al.,}{Fraine
  et~al.}{2014}]{fraineWaterVapourAbsorption2014}
Fraine J.,  et~al., 2014, \mn@doi [Nature] {10.1038/nature13785}, 513, 526

\bibitem[\protect\citeauthoryear{Hansen}{Hansen}{2019}]{hansenFormationExoplanetarySatellites}
Hansen B. M.~S.,  2019, \mn@doi [Science Advances] {10.1126/sciadv.aaw8665}, 5,
  eaaw8665

\bibitem[\protect\citeauthoryear{Heller, Rodenbeck  \& Bruno}{Heller
  et~al.}{2019}]{hellerAlternativeInterpretationExomoon2019}
Heller R.,  Rodenbeck K.,   Bruno G.,  2019, \mn@doi [Astronomy \&
  Astrophysics] {10.1051/0004-6361/201834913}, 624, A95

\bibitem[\protect\citeauthoryear{Huitson et~al.,}{Huitson
  et~al.}{2013}]{huitsonHSTOpticaltonearIRTransmission2013}
Huitson C.~M.,  et~al., 2013, \mn@doi [Monthly Notices of the Royal
  Astronomical Society] {10.1093/mnras/stt1243}, 434, 3252

\bibitem[\protect\citeauthoryear{Kass \& Raftery}{Kass \&
  Raftery}{1995}]{kassBayesFactors1995}
Kass R.~E.,  Raftery A.~E.,  1995, \mn@doi [Journal of the American Statistical
  Association] {10.1080/01621459.1995.10476572}, 90, 773

\bibitem[\protect\citeauthoryear{Kipping}{Kipping}{2011}]{kippingLunaAlgorithmGenerating2011}
Kipping D.~M.,  2011, \mn@doi [Monthly Notices of the Royal Astronomical
  Society] {10.1111/j.1365-2966.2011.19086.x}, 416, 689

\bibitem[\protect\citeauthoryear{Kipping}{Kipping}{2021}]{kippingTransitOrigamiMethod2021}
Kipping D.,  2021, \mn@doi [Monthly Notices of the Royal Astronomical Society]
  {10.1093/mnras/stab2013}, 507, 4120

\bibitem[\protect\citeauthoryear{Kipping et~al.,}{Kipping
  et~al.}{2022}]{kippingExomoonSurvey702022}
Kipping D.,  et~al., 2022, \mn@doi [Nature Astronomy]
  {10.1038/s41550-021-01539-1}, 6, 367

\bibitem[\protect\citeauthoryear{Knutson et~al.,}{Knutson
  et~al.}{2014}]{knutsonHubbleSpaceTelescope2014}
Knutson H.~A.,  et~al., 2014, \mn@doi [The Astrophysical Journal]
  {10.1088/0004-637X/794/2/155}, 794, 155

\bibitem[\protect\citeauthoryear{Konacki, Torres, Jha  \& Sasselov}{Konacki
  et~al.}{2003}]{konackiExtrasolarPlanetThat2003}
Konacki M.,  Torres G.,  Jha S.,   Sasselov D.~D.,  2003, \mn@doi [Nature]
  {10.1038/nature01379}, 421, 507

\bibitem[\protect\citeauthoryear{Kreidberg, Luger  \& Bedell}{Kreidberg
  et~al.}{2019}]{kreidbergNoEvidenceLunar2019}
Kreidberg L.,  Luger R.,   Bedell M.,  2019, \mn@doi [The Astrophysical
  Journal] {10.3847/2041-8213/ab20c8}, 877, L15

\bibitem[\protect\citeauthoryear{Martin, Fabrycky  \& Montet}{Martin
  et~al.}{2019}]{martinTransitsInclinedExomoons2019}
Martin D.~V.,  Fabrycky D.~C.,   Montet B.~T.,  2019, \mn@doi [The
  Astrophysical Journal] {10.3847/2041-8213/ab0aea}, 875, L25

\bibitem[\protect\citeauthoryear{Moraes \& Vieira~Neto}{Moraes \&
  Vieira~Neto}{2020}]{moraesExploringFormationScenarios2020a}
Moraes R.~A.,  Vieira~Neto E.,  2020, \mn@doi [Monthly Notices of the Royal
  Astronomical Society] {10.1093/mnras/staa1441}, 495, 3763

\bibitem[\protect\citeauthoryear{Pont, Knutson, Gilliland, Moutou  \&
  Charbonneau}{Pont et~al.}{2008}]{pontDetectionAtmosphericHaze2008}
Pont F.,  Knutson H.,  Gilliland R.~L.,  Moutou C.,   Charbonneau D.,  2008,
  \mn@doi [Monthly Notices of the Royal Astronomical Society]
  {10.1111/j.1365-2966.2008.12852.x}, 385, 109

\bibitem[\protect\citeauthoryear{Pontoppidan et~al.,}{Pontoppidan
  et~al.}{2016}]{pontoppidanPandeiaMultimissionExposure2016}
Pontoppidan K.~M.,  et~al., 2016, \mn@doi [Observatory Operations: Strategies,
  Processes, and Systems VI] {10.1117/12.2231768}, p.~44

\bibitem[\protect\citeauthoryear{Porter \& Grundy}{Porter \&
  Grundy}{2011}]{porterPostCaptureEvolutionPotentially2011}
Porter S.~B.,  Grundy W.~M.,  2011, \mn@doi [The Astrophysical Journal]
  {10.1088/2041-8205/736/1/L14}, 736, L14

\bibitem[\protect\citeauthoryear{Ranjan, Charbonneau, D{\'e}sert, Madhusudhan,
  Deming, Wilkins  \& Mandell}{Ranjan
  et~al.}{2014}]{ranjanAtmosphericCharacterizationFive2014}
Ranjan S.,  Charbonneau D.,  D{\'e}sert J.-M.,  Madhusudhan N.,  Deming D.,
  Wilkins A.,   Mandell A.~M.,  2014, \mn@doi [The Astrophysical Journal]
  {10.1088/0004-637X/785/2/148}, 785, 148

\bibitem[\protect\citeauthoryear{Rustamkulov, Sing, Liu  \& Wang}{Rustamkulov
  et~al.}{2022}]{rustamkulovAnalysisJWSTNIRSpec2022}
Rustamkulov Z.,  Sing D.~K.,  Liu R.,   Wang A.,  2022, \mn@doi [The
  Astrophysical Journal Letters] {10.3847/2041-8213/ac5b6f}, 928, L7

\bibitem[\protect\citeauthoryear{Sartoretti \& Schneider}{Sartoretti \&
  Schneider}{1999}]{sartorettiDetectionSatellitesExtrasolar1999}
Sartoretti P.,  Schneider J.,  1999, \mn@doi [Astronomy and Astrophysics
  Supplement Series] {10.1051/aas:1999148}, 134, 553

\bibitem[\protect\citeauthoryear{Stevenson, Bean, Seifahrt, D{\'e}sert,
  Madhusudhan, Bergmann, Kreidberg  \& Homeier}{Stevenson
  et~al.}{2014a}]{stevensonTransmissionSpectroscopyHot2014}
Stevenson K.~B.,  Bean J.~L.,  Seifahrt A.,  D{\'e}sert J.-M.,  Madhusudhan N.,
   Bergmann M.,  Kreidberg L.,   Homeier D.,  2014a, \mn@doi [The Astronomical
  Journal] {10.1088/0004-6256/147/6/161}, 147, 161

\bibitem[\protect\citeauthoryear{Stevenson, Bean, Fabrycky  \&
  Kreidberg}{Stevenson et~al.}{2014b}]{stevensonHubbleSpaceTelescope2014}
Stevenson K.~B.,  Bean J.~L.,  Fabrycky D.,   Kreidberg L.,  2014b, \mn@doi
  [The Astrophysical Journal] {10.1088/0004-637X/796/1/32}, 796, 32

\bibitem[\protect\citeauthoryear{Teachey \& Kipping}{Teachey \&
  Kipping}{2018}]{teacheyEvidenceLargeExomoon2018}
Teachey A.,  Kipping D.~M.,  2018, \mn@doi [Science Advances]
  {10.1126/sciadv.aav1784}, 4, eaav1784

\bibitem[\protect\citeauthoryear{Teachey, Kipping  \& Schmitt}{Teachey
  et~al.}{2017}]{teacheyHEKVIDearth2017}
Teachey A.,  Kipping D.~M.,   Schmitt A.~R.,  2017, \mn@doi [The Astronomical
  Journal] {10.3847/1538-3881/aa93f2}, 155, 36

\bibitem[\protect\citeauthoryear{Wakeford, Sing, Evans, Deming  \&
  Mandell}{Wakeford
  et~al.}{2016}]{wakefordMarginalisingInstrumentSystematics2016}
Wakeford H.~R.,  Sing D.~K.,  Evans T.,  Deming D.,   Mandell A.,  2016,
  \mn@doi [The Astrophysical Journal] {10.3847/0004-637X/819/1/10}, 819, 10

\makeatother
\end{thebibliography}

\appendix
\section*{Appendix}
\label{sec:appendix}
\begin{tabular}{|l|r|r|r|r|r|}
\hline
Sim \# &  Bayes Factor &  Delta $\chi^2$ &  Delta BIC &  Delta AIC & MAP\\
\hline
\rowcolor{WildStrawberry!30}0          &          0.50 &        24.33 &      13.58 &     -10.33 & F\\
\rowcolor{WildStrawberry!30}1          &          0.24 &        25.43 &      12.48 &     -11.43 & F\\
\rowcolor{WildStrawberry!30}2          &          4.03 &         5.13 &      32.78 &       8.87 & P \\
\rowcolor{Goldenrod!30}3          &       4750.15 &         0.77 &      37.15 &      13.23 & N\\
\rowcolor{WildStrawberry!30}4          &          0.29 &        17.95 &      19.96 &      -3.95 & F\\
\rowcolor{WildStrawberry!30}5          &          0.28 &         0.65 &      37.27 &      13.35 & F\\
\rowcolor{WildStrawberry!30}6          &          0.23 &         0.48 &      37.44 &      13.52 & P$^*$\\
\rowcolor{WildStrawberry!30}7          &          2.65 &         3.09 &      34.82 &      10.91  & P\\
\rowcolor{WildStrawberry!30}8          &          2.27 &        17.84 &      20.07 &      -3.84 & P\\
\rowcolor{WildStrawberry!30}9          &          0.14 &         2.04 &      35.87 &      11.96  & P\\
\rowcolor{WildStrawberry!30}10         &          0.99 &         4.55 &      33.36 &       9.45 & T\\
\rowcolor{Goldenrod!30}11         &         30.61 &         0.13 &      37.79 &      13.87 & N \\
\rowcolor{Goldenrod!30}12         &         48.77 &         2.89 &      35.02 &      11.11  & P\\
\rowcolor{ForestGreen!70}13         &         16.21 &        33.65 &       4.26 &     -19.65  & F\\
\rowcolor{WildStrawberry!30}14         &          0.44 &        10.53 &      27.38 &       3.47  & F\\
\rowcolor{WildStrawberry!30}15         &          6.21 &        25.36 &      12.55 &     -11.36  & F\\
\rowcolor{Goldenrod!30}16         &      42892.56 &         0.23 &      37.68 &      13.77 & N \\
\rowcolor{Goldenrod!30}17         &       1905.95 &         2.49 &      35.42 &      11.51  & P\\
\rowcolor{Goldenrod!30}18         &     681776.11 &         0.09 &      37.82 &      13.91  & N\\
\rowcolor{ForestGreen!30}19         &     836357.48 &        30.25 &       7.66 &     -16.25  & N\\
\rowcolor{WildStrawberry!30}20         &          2.75 &         3.66 &      34.26 &      10.34  & P\\
\rowcolor{ForestGreen!70}21         &   43803471.79 &        55.79 &     -17.88 &     -41.79  & F\\
\rowcolor{WildStrawberry!30}22         &          7.46 &         2.26 &      35.66 &      11.74  & N$^\dagger$\\
\rowcolor{ForestGreen!30}23         &         51.73 &        32.36 &       5.55 &     -18.36  & F\\
\rowcolor{WildStrawberry!30}24         &          0.50 &         8.59 &      29.33 &       5.41  & F\\
\rowcolor{Goldenrod!30}25         &        137.50 &         5.67 &      32.24 &       8.33  & P\\
\rowcolor{WildStrawberry!30}26         &          0.54 &        41.91 &      -4.00 &     -27.91  & F\\
\rowcolor{ForestGreen!30}27         &   31177616.96 &        62.20 &     -24.28 &     -48.20  & F\\
\rowcolor{Goldenrod!30}28         &        171.02 &        -0.34 &      38.25 &      14.34  & N\\
\rowcolor{WildStrawberry!30}29         &          0.28 &         1.21 &      36.70 &      12.79 & F$^*$\\
\rowcolor{WildStrawberry!30}30         &          0.26 &        16.99 &      20.93 &      -2.99  & F$^\dagger$\\
\rowcolor{WildStrawberry!30}31         &          0.21 &         0.89 &      37.02 &      13.11 & F \\
\rowcolor{Goldenrod!30}32         &         81.00 &         0.20 &      37.71 &      13.80  & N\\
\rowcolor{WildStrawberry!30}33         &          0.12 &         2.43 &      35.49 &      11.57  & F\\
\rowcolor{WildStrawberry!30}34         &          0.26 &         5.69 &      32.23 &       8.31  & F$^\dagger$\\
\rowcolor{Goldenrod!30}35         &         50.26 &         1.79 &      36.12 &      12.21  & N\\
\rowcolor{WildStrawberry!30}36         &          0.20 &         2.90 &      35.01 &      11.10  & F\\
\rowcolor{WildStrawberry!30}37         &          0.83 &         2.18 &      35.73 &      11.82  & F\\
\rowcolor{Goldenrod!30}38         &   64491383.32 &        -0.11 &      38.02 &      14.11  & N\\
\rowcolor{Goldenrod!30}39         &         11.48 &         0.15 &      37.76 &      13.85  & N\\
\rowcolor{WildStrawberry!30}40         &          0.70 &        -0.39 &      38.30 &      14.39  & N\\
\rowcolor{WildStrawberry!30}41         &          0.95 &        25.21 &      12.71 &     -11.21  & F\\
\rowcolor{ForestGreen!70}42         &        115.39 &        34.96 &       2.95 &     -20.96  & F\\
\rowcolor{WildStrawberry!30}43         &          0.12 &         1.80 &      36.12 &      12.20  & F\\
 \hline
\end{tabular}

\begin{table}
\caption{Summary statistics of each injection-recovery trial. Rows in red failed each of the 3 requirements, those in yellow satisfied the Bayes factor of 10, light green had a Bayes factor of 10 and both $\Delta \chi^2 > 9$ and $\Delta$AIC $<0$, and finally those in dark green met all statistic thresholds and correctly captured the injected moon radius and separation within the $2\sigma$ posterior bounds. The last column displays a qualitative assessment of the maximum a priori solution: F signifies a fully transiting solution, P marks a partial transit solution (the moon's ingress or egress is not captured by the baseline), and N denotes a non-transiting moon solutions. Asterisks in this column denote injected moons which themselves were non-transiting, and daggers mark injected moons whose ingress or egress were not fully captured by the baseline. In principle, moons could still be successfully recovered even when not transiting due to their influence on the planet's time of mid transit.}
\label{table:sims}
\end{table}

\begin{tabular}{|l|r|r|r|r|r|}
\hline
\rowcolor{WildStrawberry!30}44         &          5.64 &        27.83 &      10.08 &     -13.83  & F\\
\rowcolor{Goldenrod!30}45         &         30.00 &         1.90 &      36.02 &      12.10  & N\\
\rowcolor{Goldenrod!30}46         &        175.96 &         0.42 &      37.50 &      13.58  & N\\
\rowcolor{ForestGreen!70}47         &       3940.83 &        43.71 &      -5.80 &     -29.71  & F\\
\rowcolor{Goldenrod!30}48         &      28931.27 &         0.73 &      37.18 &      13.27  & N\\
\rowcolor{Goldenrod!30}49         &         12.91 &         0.89 &      37.02 &      13.11  & P\\
\rowcolor{WildStrawberry!30}50         &          1.23 &         8.95 &      28.96 &       5.05  & N\\
\rowcolor{WildStrawberry!30}51         &          0.60 &         0.53 &      37.38 &      13.47  & N\\
\rowcolor{WildStrawberry!30}52         &          0.22 &         2.50 &      35.42 &      11.50  & P\\
\rowcolor{ForestGreen!30}53         &    4018781.30 &        58.02 &     -20.10 &     -44.02  & F\\
\rowcolor{ForestGreen!70}54         &       3090.96 &        44.86 &      -6.95 &     -30.86  & F\\
\rowcolor{Goldenrod!30}55         &         19.72 &         4.47 &      33.45 &       9.53  & P\\
\rowcolor{WildStrawberry!30}56         &          0.31 &         2.88 &      35.03 &      11.12  & F\\
\rowcolor{WildStrawberry!30}57         &          0.56 &        22.88 &      15.03 &      -8.88  & F\\
\rowcolor{WildStrawberry!30}58         &          0.31 &         1.57 &      36.34 &      12.43  & F\\
\rowcolor{ForestGreen!70}59         &        146.22 &        38.55 &      -0.63 &     -24.55  & F\\
\rowcolor{WildStrawberry!30}60         &          0.32 &         2.47 &      35.45 &      11.53  & N\\
\rowcolor{WildStrawberry!30}61         &          0.14 &         5.02 &      32.89 &       8.98  & F\\
\rowcolor{WildStrawberry!30}62         &          0.11 &        16.82 &      21.09 &      -2.82  & F\\
\rowcolor{WildStrawberry!30}63         &          2.93 &         0.39 &      37.52 &      13.61  & N\\
\rowcolor{WildStrawberry!30}64         &          0.97 &        19.81 &      18.10 &      -5.81  & P$^\dagger$\\
\rowcolor{WildStrawberry!30}65         &          0.38 &        -0.17 &      38.08 &      14.17  & N$^*$\\
\rowcolor{WildStrawberry!30}66         &          0.17 &         5.68 &      32.23 &       8.32  & F\\
\rowcolor{WildStrawberry!30}67         &          0.82 &         1.21 &      36.70 &      12.79  & F\\
\rowcolor{Goldenrod!30}68         &     114158.66 &         8.93 &      28.99 &       5.07  & P\\
\rowcolor{WildStrawberry!30}69         &          0.27 &        17.68 &      20.23 &      -3.68  & F\\
\rowcolor{ForestGreen!70}70         &   73656610.69 &        57.15 &     -19.24 &     -43.15  & F\\
\rowcolor{WildStrawberry!30}71         &          1.13 &         9.74 &      28.17 &       4.26  & P\\
\rowcolor{WildStrawberry!30}72         &          9.48 &         1.41 &      36.51 &      12.59  & N\\
\rowcolor{WildStrawberry!30}73         &          1.37 &         0.53 &      37.38 &      13.47  & N\\
\rowcolor{WildStrawberry!30}74         &          0.59 &        26.44 &      11.48 &     -12.44  & F\\
\hline
\end{tabular}

%%%%%%%%%%%%%%%%%%%%%%%%%%%%%%%%%%%%%%%%%%%%%%%%%%

% Don't change these lines
\bsp	% typesetting comment
\label{lastpage}
\end{document}